\documentclass[aps,superscriptaddress,showpacs,showkeys,nofootinbib,twocolumn,floatfix]{revtex4}
\usepackage{graphicx,epsfig,amssymb}
\usepackage{color}
\usepackage{amsfonts}
\usepackage{amsmath}
\usepackage{latexsym}      
\definecolor{darkgreen}{rgb}{0,0.65,0}

\newcommand{\be}{\begin{equation}}
\newcommand{\ee}{\end{equation}}
\newcommand{\ba}{\begin{eqnarray}}
\newcommand{\ea}{\end{eqnarray}}
\newcommand{\la}{\langle}
\newcommand{\ra}{\rangle}
\newcommand{\di}{ {\rm d} }

\newcommand{\tvec}[1]{\mbox{\boldmath{$#1$}}}

\begin{document}
\newcommand*{\Pavia}{Dipartimento di Fisica Nucleare e Teorica, 
    Universit\`a degli Studi di Pavia, Pavia, Italy}\affiliation{\Pavia}
\newcommand*{\INFN}{Istituto Nazionale di Fisica Nucleare, 
    Sezione di Pavia, Pavia, Italy}\affiliation{\INFN}
\newcommand*{\UConn}{Department of Physics, University of Connecticut, 
    Storrs, CT 06269, USA}\affiliation{\UConn}

\title{
    Naive time-reversal odd phenomena in 
    semi-inclusive deep-inelastic scattering\\
    from light-cone constituent quark models}

\author{B.~Pasquini}\affiliation{\Pavia}\affiliation{\INFN}
\author{P.~Schweitzer}\affiliation{\UConn}

\date{March 2011}
\begin{abstract}
  We present results for leading-twist azimuthal asymmetries in 
  semi-inclusive lepton-nucleon deep-inelastic scattering due to 
  naively time-reversal odd transverse-momentum dependent parton 
  distribution functions from the light-cone constituent quark 
  model.
  We carefully discuss the range of applicability of the model, 
  especially with regard to positivity constraints and evolution
  effects.
  We find good agreement with available experimental data 
  from COMPASS and HERMES, and present predictions to be tested 
  in forthcoming experiments at Jefferson Lab.
\end{abstract}
\pacs{13.88.+e, 
      13.85.Ni, 
      13.60.-r, 
      13.85.Qk} 
\keywords{semi-inclusive deep inelastic scattering,
      transverse-momentum dependent distribution functions}
\maketitle


\section{Introduction}
\label{Sec-1:introduction}

\begin{figure}

        \vspace{15mm} 

        \includegraphics[width=8cm]{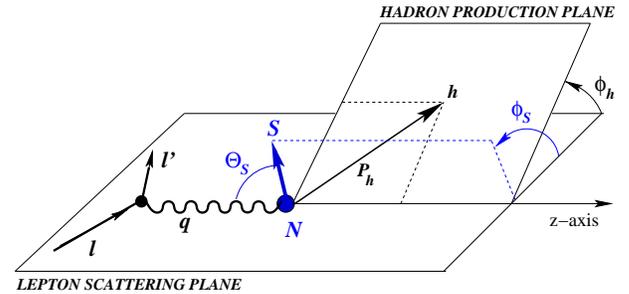}
        \caption{\label{Fig1-SIDIS-kinematics}
    	Kinematics of the SIDIS process $lN\to l^\prime h X$.
        The azimuthal angles of produced hadron 
        and nucleon polarization vector are $\phi_h$ and $\phi_S$.
        The transverse momentum of the hadron is 
        $P_{h\perp}\ll Q$ where $Q^2=-q^2=(l-l^\prime)^2$.}
\end{figure}

Two out of the 18 structure functions describing the semi-inclusive 
lepton-nucleon deep-inelastic scattering (SIDIS) process 
\cite{Kotzinian:1994dv,Mulders:1995dh}, see Fig.~\ref{Fig1-SIDIS-kinematics},
are associated at leading order of the hard scale $Q$ with 
{\sl naively time-reversal odd (T-odd)} transverse-momentum dependent 
parton distributions (TMDs), the Sivers \cite{Sivers:1989cc} and 
Boer-Mulders  \cite{Boer:1997nt} functions.
Their existence is ultimately related to initial and final state 
interactions in QCD \cite{Brodsky:2002cx} encoded in appropriately 
defined Wilson lines \cite{Collins:2002kn,Belitsky:2002sm}. 
T-odd TMDs have unusual ``universality'' properties, and are 
predicted to have opposite signs \cite{Collins:2002kn} in SIDIS 
and the Drell-Yan process. The basis for this description is a
generalized factorization approach which applies 
when the final state transverse momentum is small compared to the 
hard scale \cite{Collins:1981uk,Ji:2004wu,Collins:2004nx}, i.e.\  
$P_{h\perp}\ll Q$ in SIDIS. 

Data on the Sivers Boer-Mulders effect from SIDIS are available
or forthcoming
\cite{Airapetian:2004tw,Alexakhin:2005iw,Alekseev:2008dn,Alekseev:2010rw,
Airapetian:2009ti,Mkrtchyan:2007sr,Osipenko:2008rv,Giordano:2010gq,Sbrizzai,
Gohn:2009zz,Hall-A-neutron,Gao:2010av,Avakian:LOI,Anselmino:2011ay}
(for a recent review, see Ref.~\cite{Barone:2010zz}).
Both effects were subject to phenomenological studies in SIDIS
\cite{Efremov:2004tp,Anselmino:2005nn,Vogelsang:2005cs,Collins:2005ie,
Barone:2005kt,Anselmino:2008sg,Barone:2009hw,Bacchetta:2010si}
and DY
\cite{Boer:1999mm,Collins:2005rq,Bianconi:2005yj,
Sissakian:2008th,Kang:2009sm,Lu:2009ip}.
General aspects of the Sivers and Boer-Mulders functions were discussed in
\cite{Bacchetta:1999kz,Pobylitsa:2003ty,Burkardt:2004ur,Brodsky:2006hj},
and quark model calculations were reported in 
\cite{Brodsky:2002cx,Yuan:2003wk,Gamberg:2003ey,Meissner:2007rx,
Gamberg:2007wm,Courtoy:2008vi,Bacchetta:2008af,Kotzinian:2008fe,
Ellis:2008in,Gamberg:2009uk,Pasquini:2010af}.
Common to the quark model approaches is that the final (in SIDIS) 
or initial (in DY) state interactions are modeled by means of a 
one-gluon exchange (in \cite{Gamberg:2009uk} steps were made 
to go beyond that).

Among the most recent studies are the calculations in the 
light-cone constituent quark model (LCCQM) \cite{Pasquini:2010af}.
On the basis of the one-gluon-exchange approximation formulated 
in the light-cone quantization formalism \cite{Brodsky:2010vs} 
the Sivers and Boer-Mulders were modeled in \cite{Pasquini:2010af} 
using light-cone wave-functions which were previously applied 
with success to calculations of transversity, electroweak properties 
of the nucleon, generalized parton distributions, distribution 
amplitudes~\cite{Pasquini:2005dk,Pasquini:2007iz,Pasquini:2009ki}, 
and T-even leading-twist TMDs \cite{Pasquini:2008ax}.
The results illustrate the relevance of different orbital angular 
momentum components of the nucleon light-cone wave function.
 
%
%

The T-even TMDs from this model \cite{Pasquini:2008ax} were shown
to yield results in satisfactory agreement with available SIDIS data 
\cite{Boffi:2009sh}. The purpose of this work is to explore whether 
also the T-odd TMDs from this approach \cite{Pasquini:2010af} can
explain the corresponding SIDIS data. It is important to consider that 
in contrast to previous studies \cite{Pasquini:2005dk,
Pasquini:2007iz,Pasquini:2009ki,Pasquini:2008ax}, 
we here probe more than the modeled nucleon light-cone wave-functions.
In the present work we moreover also probe to which extent 
the ``one-gluon-exchange approximation'' for T-odd TMDs works.

The article is organized as follows. 
In Sec.~\ref{Sec-2:SIDIS} the SIDIS process, 
and relevant observables are briefly presented.
In Sec.~\ref{Sec-3:model} the calculation of the T-odd TMDs 
\cite{Pasquini:2010af} is reviewed, positivity and evolution
effects are discussed, and limitations of the approach
conservatively disclosed.
In Secs.~\ref{Sec-4:AUT} and \ref{Sec-5:AUU} we discuss the 
numerical results for the asymmetries and compare them to SIDIS 
data. Conclusions are drawn in Sec.~\ref{sect:conclusions}.
The appendix contains some remarks about the corrections 
due to the Cahn effect in the 
$\cos(2\phi_h)$-asymmetry in unpolarized SIDIS.

\section{Sivers and Boer-Mulders in SIDIS}
\label{Sec-2:SIDIS}

The kinematics of the SIDIS process, the momenta $l,\,l'$,
$q$, $Q^2$ and $P_{h\perp}=|{\tvec P}_{\!h\perp}|$ are defined in
Fig.~\ref{Fig1-SIDIS-kinematics}. The SIDIS variables are 
$x = Q^2/(2P\cdot q)$, $y = (P\cdot q)/(P\cdot l)$, and
$z = (P\cdot P_h)/(P\cdot q)$ where $P$ is the nucleon momentum.
Denoting by $\sigma_0$ its spin- and $\phi_h$-independent part, 
the SIDIS cross-section $\sigma$ 
(differential in $x$, $y$, $z$ and the azimuthal angle $\phi_h$ 
which we do not indicate for brevity)
can be written as
\ba
    \di^4\sigma
     =	\di^4\sigma_0
  \Biggl\{ 1+\cos(2\phi_h)\,p_1(y)\,A_{UU}^{\cos(2\phi_h)}&&\nonumber\\
     +    S_T\sin(\phi_h-\phi_S)\,  A_{UT}^{\sin( \phi_h-\phi_S)}&& 
     \hspace{-2mm}+\dots \Biggr\} \;\;\; \label{Eq:azim-distr-in-SIDIS}
\ea
where $p_1(y) = (1-y)/(1-y+\frac12\,y^2)$ up to (systematically
neglected) power-suppressed
terms, and the dots indicate terms due to T-even TMDs or subleading-twist
\cite{Bacchetta:2006tn}. 

$A_{UU}^{\cos(2\phi_h)}=F_{UU}^{\cos(2\phi_h)}/F_{UU}$ and similarly
$A_{UT}^{\sin(\phi_h-\phi_S)}$ are the azimuthal asymmetries 
defined in terms of structure functions. $F_{UU}$ is the 
unpolarized structure function which gives rise to $\sigma_0$. 
Here the first index U denotes the unpolarized electron beam. 
The second index U/T refers to the target polarization which can be 
unpolarized/transverse with respect to the virtual photon
(in experiment the transverse target polarization is with respect to the 
beam, of course, but this is up to power-corrections the same).
The superscript reminds us of the kind of angular distribution of the 
produced hadrons with no index describing an isotropic $\phi_h$-distribution.
In the Bjorken-limit the relevant structure functions are given  
at tree-level by \cite{Bacchetta:2006tn}
\ba
\hspace{-7mm}
&& F_{UU} =\phantom{-}\,{\cal C}\biggl[\;f_1 D_1 \;\biggr],
        {\phantom{\Biggl|}} \label{Eq:FUU}\\
\hspace{-7mm}
&& F_{UT}^{\sin\left(\phi_h -\phi_S\right)} = -\,{\cal C}\biggl[
	\frac{{\tvec{\hat{h}}}\cdot {\tvec p}_T^{ }}{M}f_{1T}^{\perp}D_1\biggr],
	\label{Eq:FUTSiv}\\
\hspace{-7mm}
&& F_{UU}^{\cos(2\phi_h)}={\cal C}\biggl[
   	\frac{2\, \bigl({\tvec{\hat{h}}} \cdot {\tvec K}_{\!T}^{ } \bigr)
   	\,\bigl({\tvec{\hat{h}}} \cdot {\tvec p}_T^{ } \bigr)
    	-{\tvec K}_{\!T}^{ }\cdot{\tvec p}_T^{ }}{z\,m_h M}
    	h_{1}^{\perp} H_{1}^{\perp }\biggr],\;	\label{Eq:FUUcos2phi}
\ea
where ${\tvec{\hat{h}}}={\tvec P}_{h\perp}/P_{h\perp}$ and $M$ ($m_h$) is 
the mass of nucleon (produced hadron). The convolutions are defined as 
\ba
&& \hspace{-7mm} {\cal C}\biggl[\;w\;j\;J\biggr]=
   \int\di^2{\tvec p}_T^{ }\int\di^2{\tvec K}_T
   \;\delta^{(2)}(z\,{\tvec p}_T^{ }+{\tvec K}_T^{ }-{\tvec P}_{h\perp}^{ })
   \nonumber\\
   &&\times \,w({\tvec p}_T,\,{\tvec K}_T)\,
   \sum_a e_a^2\;x\,j^a(x,p_T)\;J^a(z, K_T),\label{Eq:conv-integral}
\ea
with a generic TMD $j^a$ and transverse momentum dependent fragmentation 
function $J^a$, and $p_T=|{\tvec p}_T|$ and $K_T=|{\tvec K}_T|$. 
In Eqs.~(\ref{Eq:FUU}--\ref{Eq:FUUcos2phi}) $D_1^a$ 
is the unpolarized and $H_1^{\perp a}$ the Collins 
\cite{Collins:1992kk,Efremov:1992pe}
fragmentation function.
Notice that in this tree-level treatment one neglects 
soft factors \cite{Collins:1981uk,Ji:2004wu,Collins:2004nx}.

In order to solve the convolution integrals we will make use of 
the Gaussian Ansatz. This step could, in principle, be avoided by
considering adequately weighted asymmetries \cite{Boer:1997nt}.
It is given by
\ba
&&  j^a(x,p_T) = j^a(x)\;
    \frac{\exp(-p_T^{\:2}/\la p_T^2(j)\ra)}{\pi\la p_T^2(j)\ra} \;,\nonumber\\
&&  J^a(z,K_T) = J^a(z)\;
    \frac{\exp(-K_T^2/\la K_T^2(J)\ra)}{\pi\;\la K_T^2(J)\ra}\;.
    \label{Eq:Gauss-ansatz}\ea
Independently of the model for transverse momenta
$F_{UU}(x,z)=\sum_a e_a^2 \,x\,f_1^a(x)\,D_1^a(z)$,
while with the Ansatz (\ref{Eq:Gauss-ansatz}) 
we obtain for the structure functions \cite{Boffi:2009sh}
\ba
&& \hspace{-7mm} F_{UT}^{\sin\left(\phi_h -\phi_S\right)}(x,z) 
   = - B_0\sum_a e_a^2 \,x\,f_{1T}^{\perp(1)a}(x)\,D_1^a(z),
   \label{Eq:GaussFUTSiv}\\ 
&& \hspace{-7mm} F_{UU}^{\cos(2\phi_h)}(x,z)    
   = B_2\sum_a e_a^2\,x\,h_{1}^{\perp(1)a}(x)H_1^{\perp(1/2)a}(z),
   \label{Eq:GaussFUUcos2phiBM}\\
&& B_0=\frac{z\;\sqrt{\pi}\,M}
   {\{z^2\,\la p_T^2(f_{1T}^\perp)\ra+\la K_T^2(D_1)\ra\}^{1/2}},\\
&& B_2 =\frac{8\,z\,M\, [\pi\la K_T^2(H_1^\perp)\ra]^{-1/2}}
   {1+ z^2\la p_T^2(h_1^\perp)\ra \,/\, \la K_T^2(H_1)\ra },    
\ea
with the transverse moments defined as
\ba
&&   j^{\perp(1)a}(x)=\int\di^2{\tvec p}_T^{ }\,\frac{p_T^2}{2M^2}\,
     j^{\perp   a}(x,p_T)\,, \nonumber\\
&&   J^{\perp(1/2)a}(z)=\int{\rm d}^2{\tvec K}_T^{ }\,\frac{K_T}{2zm_h}
     J^{\perp     a}(z,K_T)\,.
   \label{Eq:(1)-moments}
\ea
The Gaussian Ansatz (\ref{Eq:Gauss-ansatz}) is for $f_1^a$ and $D_1^a$ 
phenomenologically well supported for $\la P_{h\perp}\ra\ll Q$
\cite{D'Alesio:2004up,Schweitzer:2010tt}. 
For polarized TMDs it is supported approximately by some models 
\cite{Pasquini:2008ax,Boffi:2009sh,Avakian:2010br,Efremov:2010mt}.
A rigorous description of $p_T$-effects would require methods 
along the QCD-based formalism of \cite{Collins:1984kg},
as implemented in \cite{Landry:2002ix,Aybat:2011zv},
but for our purposes the Ansatz (\ref{Eq:Gauss-ansatz}) 
will be sufficient.

It is important to stress  that the description of the SIDIS 
process in the Bjorken-limit within TMD factorization is generically 
valid only up to power corrections suppressed by the large scale $Q$ 
\cite{Ji:2004wu,Collins:2004nx}. 
Typically such corrections do not factorize 
(TMD factorization beyond leading twist is presently unclear 
\cite{Gamberg:2006ru} in SIDIS)
and must be excluded 
experimentally by studying the $Q$-behavior of the observables. 
In the case of $F_{UU}^{\cos2\phi_h}$ the parton model provides a way 
to estimate one of the possible power corrections which is due to 
the Cahn-effect \cite{Cahn:1978se}. This effect gives rise to 
$\cos\phi_h$ \cite{Anselmino:2005nn,Schweitzer:2010tt} and
$\cos(2\phi_h)$ modulations in the unpolarized SIDIS cross section
which are suppressed by respectively $\la p_T(f_1)\ra/Q$ and 
$\la p_T^2(f_1)\ra/Q^2$. In particular, the latter contributes
a power-correction to the Boer-Mulders asymmetry, which
is not negligible in the kinematics of present SIDIS 
experiments \cite{Giordano:2010gq,Sbrizzai}.

The Cahn-effect and its power correction to $A_{UU}^{\cos(2\phi_h)}$,
however, are to a good approximation flavor independent 
\cite{Barone:2009hw,Schweitzer:2010tt}. Below in
Sec.~\ref{Sec-5:AUU} we will therefore consider 
the difference of the asymmetries for $\pi^-$ and $\pi^+$.
Further details are discussed in the Appendix.

\section{T-odd TMDs in the LCCQM}
\label{Sec-3:model}

In this section we briefly review the calculation of the T-odd TMDs 
in the light-cone constituent quark model \cite{Pasquini:2010af}.
Within this model, the gauge-link operator entering the quark 
correlation function
which defines the TMDs is approximated by taking into account 
one-gluon exchange between the struck quark and the spectator 
quarks in the nucleon described by (real) light-cone wave functions 
(LCWFs). Working in the light-cone gauge $A^+=0$, 
the T-odd quark TMDs are written in terms
of overlap of LCWFs, convoluted with 
the gluon propagator obtained from the expansion of the gauge link. 
The final result is of order $\alpha_s$, which
enters as an overall multiplicative factor and is understood 
at the initial scale $\mu_0$ of the model.

The value of $\alpha_s(\mu_0)$ was fixed by determining the initial 
model scale $\mu_0$ following \cite{Pasquini:2004gc}
as follows. Since the model has only valence quark
degrees of freedom, one requires 
$\mu_0$ to be such that evolving the (total in the model) 
momentum fraction carried by valence quarks, 
$\la x(\mu_0)\ra_V=\sum_q\int{\rm d}x\,x(f_1^q-f_1^{\bar q})(x,\mu_0)=1$,
from $\mu_0$ to experimentally relevant scales one
matches the NLO phenomenological value 
$\la x(Q)\ra_V = 0.36$ 
at $Q^2=10\,{\rm GeV}^2$ \cite{Martin:2009iq}.
In previous works $\mu_0$ was tuned to reproduce 
exactly the phenomenological value of $\la x(Q)\ra_V$.
Here we content ourselves with reproducing it 
within 10$\,\%$, which is acceptable
in view of the generic model accuracy of  (10--30)$\,\%$ 
\cite{Boffi:2009sh}.
Allowing for a $10\,\%$-overestimate of $\la x(Q)\ra_V$ yields a 
higher $\mu_0$ and smaller $\alpha_s(\mu_0)$, and 
better convergence. 
The evolution was performed in the variable flavor-number scheme with 
heavy-quark mass thresholds $m_c=1.4$ GeV, $m_b=4.75$ GeV, $m_t=175$ GeV 
with $\alpha_s^{{\rm NLO}}(M_Z^2)=0.12018$ at next-to-leading order (NLO,
$\overline{\mbox{\footnotesize{MS}}}$ scheme) from \cite{Martin:2009iq}. 
We obtain for the initial model scale 
\ba
   \mu_0^{\rm NLO}= 508\, \mbox{MeV}, && 
   \frac{\alpha_s^{\rm NLO}(\mu_0^2)}{4\pi}=0.128
\label{eq:alpha-nlo}
\ea
(notice that 
$\Lambda_{\rm{NLO}}^{(3,4,5)}=$ 402, 341, 239 MeV in \cite{Martin:2009iq}).
The strategy of fixing the model scale is basically the same as in 
previous works. Here we introduced the variable flavor number scheme,
and updated the value of $\alpha_s(M_Z^2)$. 
As a result, we find a higher value for the hadronic scale of the 
model, corresponding to a somewhat smaller coupling compared to 
\cite{Pasquini:2010af}.
For the calculations of T-odd TMDs, we will adopt the value 
of $\alpha_s$ in Eq.~(\ref{eq:alpha-nlo}).

The LCWF of the nucleon is modeled as described in 
\cite{Pasquini:2010af,Pasquini:2008ax}.
In particular, to disentangle the spin-spin and spin-orbit quark correlations 
encoded in the TMDs, we expand the three-quark LCWF in a basis of eigenstates 
of orbital angular momentum, which yields 6 independent 
amplitudes corresponding to different combinations of quark helicity 
and orbital angular momentum~\cite{Ji:2002xn}. 

Assuming SU(6) symmetry, these light-cone amplitudes have 
a particularly simple structure, with spin and isospin dependence 
factorized from a momentum-dependent function with spherical symmetry.
Under this assumption the orbital angular momentum content of the wave 
function is fully generated by the Melosh rotations  
which boost the rest-frame (canonical) spin 
of the quarks into light-cone helicities. 
For the momentum-dependent part of the LCWF we adopt the phenomenological 
description with parameters fitted to hadronic structure constants from
\cite{Schlumpf:1992ce} which gave satisfactory results 
in previous works \cite{Pasquini:2007iz,Pasquini:2009ki,
Pasquini:2008ax,Boffi:2009sh}.

\begin{figure}[t]
 	\centering
\includegraphics[width=7cm]{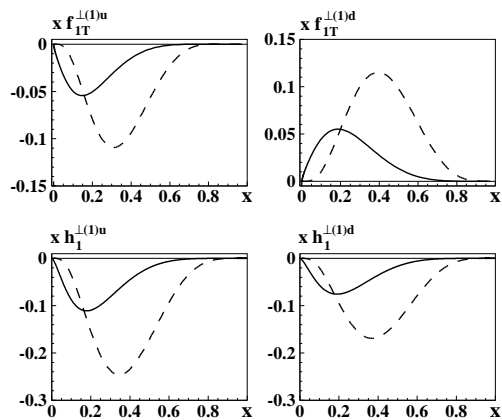}
	\caption{\label{Fig2:TMD}
    Transverse moments $f_{1T}^{\perp(1)q}(x)$ and $h_{1}^{\perp(1)q}(x)$
    in a proton for up (left) and down (right) quark, as function of $x$. 
    The dashed curves show the results at the hadronic scale $\mu_0$ of the model . The solid curves correspond to the results 
    after LO evolution to $Q^2=2.5$ GeV$^2$, using the evolution patterns 
    of the unpolarized parton distribution (transversity) for the Sivers 
    (Boer-Mulders) function.}
\end{figure}

The results for the (1)-moments of the Sivers and Boer-Mulders functions 
as defined in (\ref{Eq:(1)-moments}) are shown in Fig.~\ref{Fig2:TMD}.
The dashed curves show the results at the hadronic scale of the model, 
while the solid curves are obtained by applying LO evolution
to $Q^2=2.5$ GeV$^2$~\cite{footnote-1}.
Since the exact evolution equations for the T-odd TMDs
are still under study~\cite{Ceccopieri:2005zz,Cherednikov:2009wk,
Kang:2008ey,Zhou:2008mz,Braun:2009vc}, we use those 
evolution equations which seem most promising to ``simulate'' 
the correct evolution.

The chiral-even Sivers function is evolved according to 
the LO {\sl non-singlet} evolution pattern of $f_1^a(x)$,
which has the advantage of preserving 
the Burkardt sum rule~\cite{Burkardt:2004ur}.
This non-trivial constraint for model calculations is satisfied 
in the LCCQM at the initial model scale \cite{Pasquini:2010af}.
For the chiral-odd Boer-Mulders function we use the evolution 
pattern of $h_1^a(x)$.
The so evolved model results are consistent with the first 
extractions 
\cite{Efremov:2004tp,Anselmino:2005nn,Vogelsang:2005cs,
Anselmino:2008sg,Collins:2005ie,Barone:2009hw}
concerning signs and  magnitudes of the various flavors,
and the positions of their maxima in $x$.

%
%

As can be seen in Fig.~\ref{Fig2:TMD}, evolution effects 
are important, and this raises the question how reliably we 
estimated them here. The exact answer to this question will 
be given after the TMD-evolution formalism discussed in 
\cite{Aybat:2011zv} for $f_1^a(x,p_T)$ has been generalized 
to include the Sivers and Boer-Mulders effects. 
Until then, we can try to gain some intuition by 
investigating how much the results would change if we used 
different ways of simulating the evolution effects for T-odd 
TMDs.
For that we explored also alternative evolution patterns, and 
found that the corresponding results vary within (10--20)$\,\%$. 
This indicates that  the theoretical uncertainty associated with 
evolution effects is not larger than the model accuracy~\cite{footnote-2}.
An interesting result is that the model supports approximately
the Gaussian Ansatz, both for T-odd \cite{Pasquini:2010af} and
T-even \cite{Boffi:2009sh} TMDs.

A point deserving particular attention are inequalities among 
twist-2 TMDs \cite{Bacchetta:1999kz}, which can be written as 
\ba
&& \hspace{-7mm} {\cal P}_{\rm Siv}^q(x,p_T) \equiv 
      f_1^q(x,p_T)^2-g_1^q(x,p_T)^2 \nonumber\\ 
&&  - \frac{p_T^2}{M_N^2}\; g_{1T}^{\perp q}(x,p_T)^2 
    - \frac{p_T^2}{M_N^2}\; f_{1T}^{\perp q}(x,p_T)^2 \ge 0   
    \;,\;\; \label{Ineq:f1Tperp}\\
&& \hspace{-7mm}  {\cal P}_{\rm BM}^q(x,p_T) \equiv 
      f_1^q(x,p_T)^2-g_1^q(x,p_T)^2 \nonumber\\ 
&&  - \frac{p_T^2}{M_N^2}\; h_{1L}^{\perp q}(x,p_T)^2 
    - \frac{p_T^2}{M_N^2}\; h_1   ^{\perp q}(x,p_T)^2 \ge 0
    \;.\;\; \label{Ineq:h1perp}
\ea
In a consistent model framework the inequalities 
(\ref{Ineq:f1Tperp})-(\ref{Ineq:h1perp}) should hold. 
However, as observed in \cite{Kotzinian:2008fe}, for certain 
values of $x$ and $p_T$ numerous models violate already the 
weaker inequalities 
$\frac{p_T}{M_N}\,|f_{1T}^{\perp q}(x,p_T)| \le f_1^q(x,p_T)$, and
$\frac{p_T}{M_N}\,|h_1^{\perp q}(x,p_T)| \le f_1^q(x,p_T)$
following from (\ref{Ineq:f1Tperp}), (\ref{Ineq:h1perp}).

The reason is apparent. 
The calculations of T-odd TMDs are within the given models correct. 
But while T-odd TMDs are evaluated to ${\cal O}(\alpha_s)$, the 
``expansion'' of T-even TMDs is truncated in the quark models at 
${\cal O}(\alpha_s^0)$. In order to preserve unitarity,
and hence the inequalities, one should evaluate also
T-even TMDs consistently to order ${\cal O}(\alpha_s)$.
To best of our knowledge, this has so far not been done
in any model, and would also go beyond the scope of the 
present work.

\begin{figure}[t!]
\includegraphics[width=8cm]{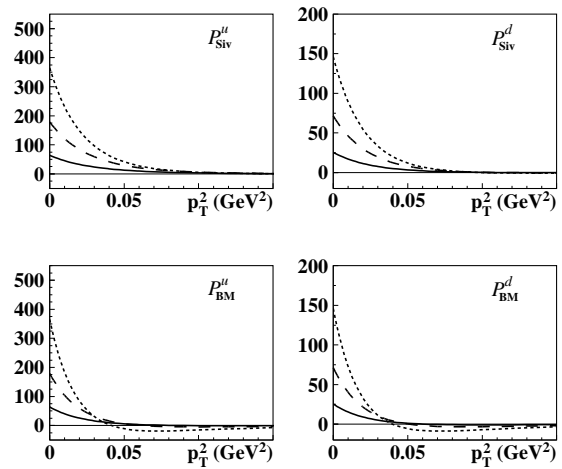}

\vspace{-5mm}

\caption{\label{Fig3:positivity}
The positivity relations involving the Sivers (upper panels) and 
Boer-Mulders functions (lower panels) 
from Eqs.~(\ref{Ineq:f1Tperp}) and (\ref{Ineq:h1perp}), respectively, 
as function of $p^2_T$ at different values of $x$: 
 $x=0.3$ (short-dashed curves), $x=0.4$ (long-dashed curves), 
and $x=0.5$ (solid curves). 
The left (right) panels show the results for up (down) quark.}
\end{figure}

Having established that one cannot expect the inequalities
(\ref{Ineq:f1Tperp},~\ref{Ineq:h1perp}) to be satisfied in 
our approach either, let us see for which values of
$x$ and $p_T$ they are violated.
For that,  in Fig.~\ref{Fig3:positivity} we plot ${\cal P}^q_{\rm Siv,BM}(x,p_T)$,
Eqs.~(\ref{Ineq:f1Tperp},~\ref{Ineq:h1perp}), vs.\ $p_T^2$ 
for selected values of $x=0.2$, $0.3$, $0.4$ at the
low model scale.
${\cal P}^q_{\rm Siv,BM}(x,p_T)$ should be always positive,
and we see that this condition is violated only at small
$x$ and for $p_T^2$ significantly larger than the respective 
$\la p_T^2\ra$. Now it is important to recall several facts.
First, quark models do not describe reliably small $x$.
Notice that the region of $x\lesssim 0.2$ at the low model scale
corresponds after (correct or approximate) evolution to 
$x\lesssim{\cal O}(10^{-2})$. Thus, practical consequences
(if any) of the violation of inequalities are beyond the
region of $x$ where quark models can be applied.
Second, in the following we will restrict ourselves to 
the use of $p_T$-integrated results for the (1)-moments,
and those are dominated by the regions of $p_T$ of the 
order of magnitude of $\la p_T^2\ra$ and smaller, i.e.\ 
where the inequalities are satisfied.

%
%

To conclude, though we have to admit that the model
predictions do not satisfy the inequalities for all $x$ and $p_T$,
we can be assured that --- in the way we will use them
for phenomenological calculations --- this will 
not yield any artifacts. In this sense we will consider
our results as compliant with positivity constraints.

\section{\boldmath Sivers asymmetry}
\label{Sec-4:AUT}

%
\begin{figure*}[t!]
  \includegraphics[height=5.2cm]{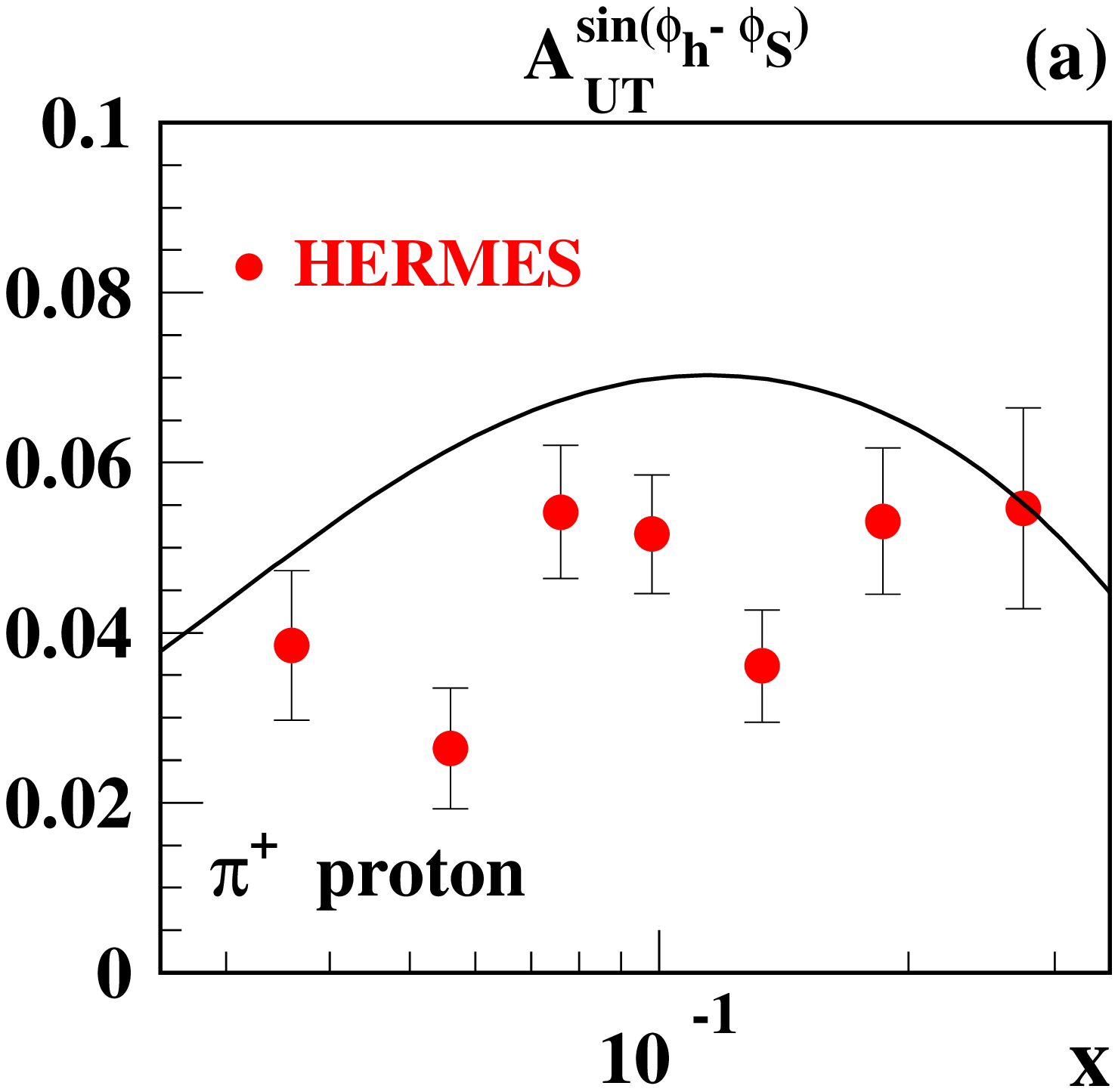}
\hspace{-8mm}
  \includegraphics[height=5.2cm]{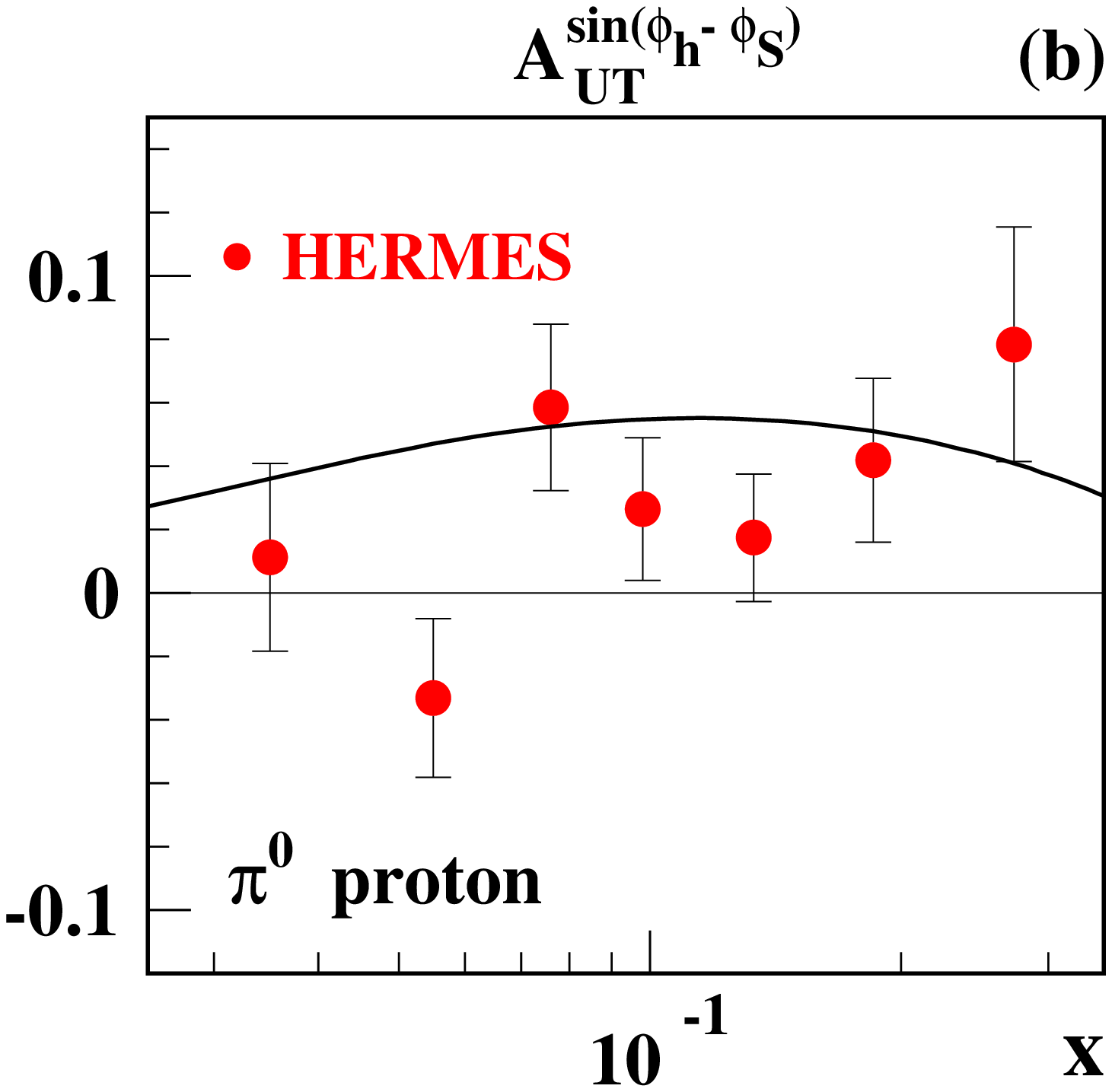}
\hspace{-8mm}
  \includegraphics[height=5.2cm]{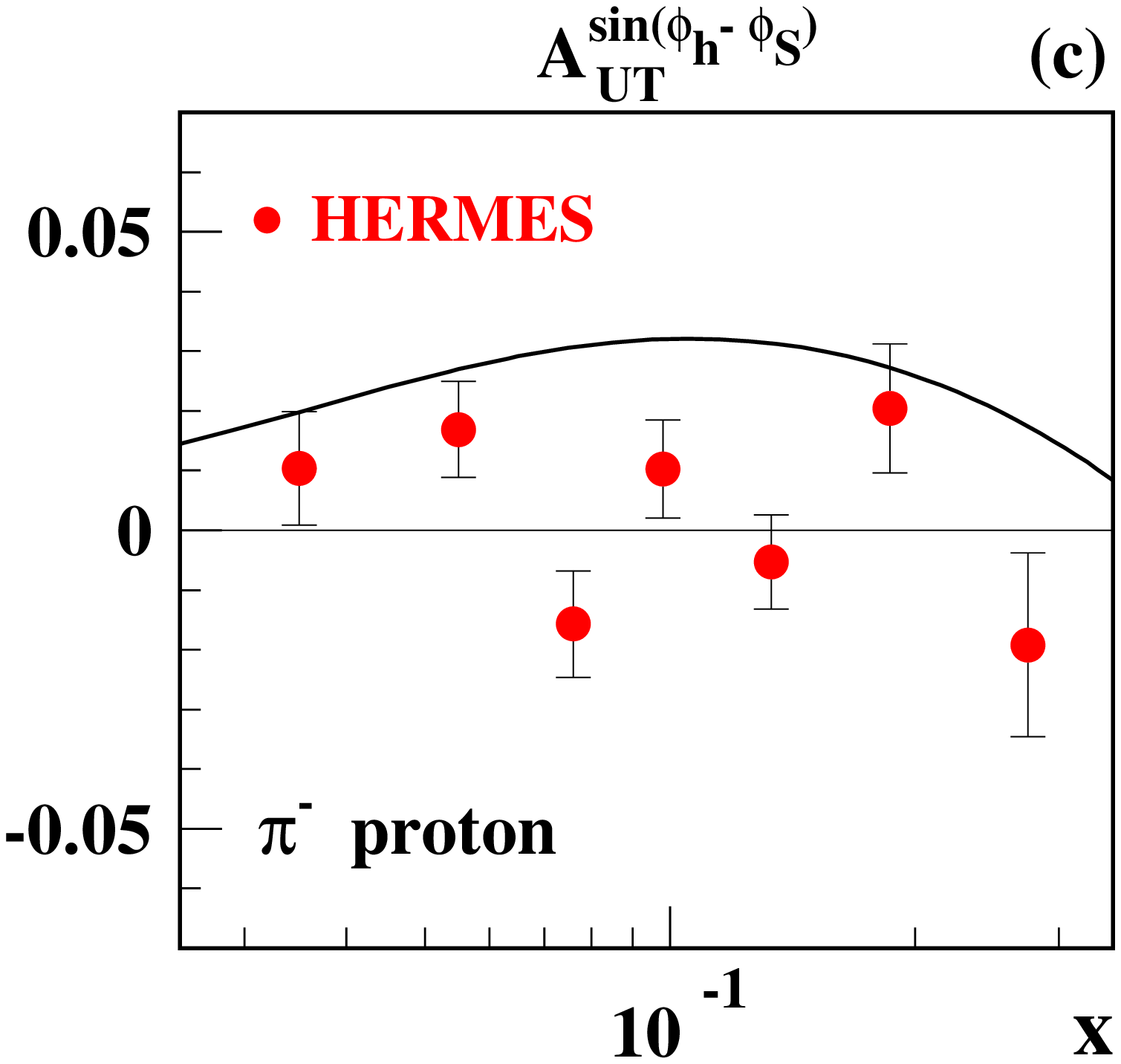}

\vspace{-9mm}

	\caption{\label{Fig-4:Sivers-HERMES}
	The single-spin asymmetry $A_{UT}^{\sin(\phi_h-\phi_S)}$ for 
        pion production off proton in SIDIS, as function of $x$. 
        The HERMES data are from \cite{Airapetian:2009ti}.
        The theoretical curves are obtained on the basis of the 
        LCCQM predictions for $f_{1T}^{\perp(1)q}(x)$ \cite{Pasquini:2010af}.}

\hspace{-2mm}
\includegraphics[height=5.2cm]{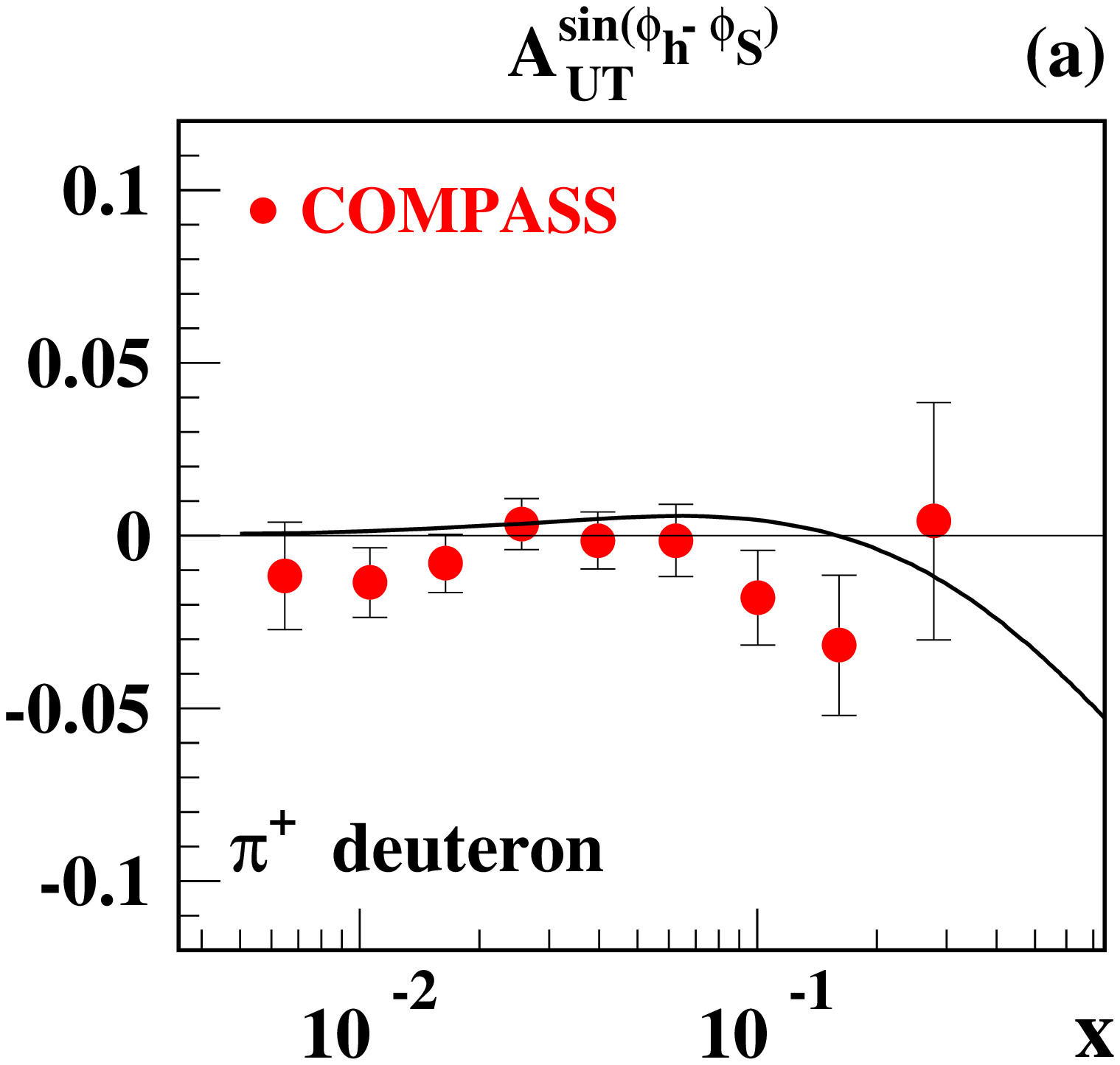}
\hspace{-12mm}
\includegraphics[height=5.2cm]{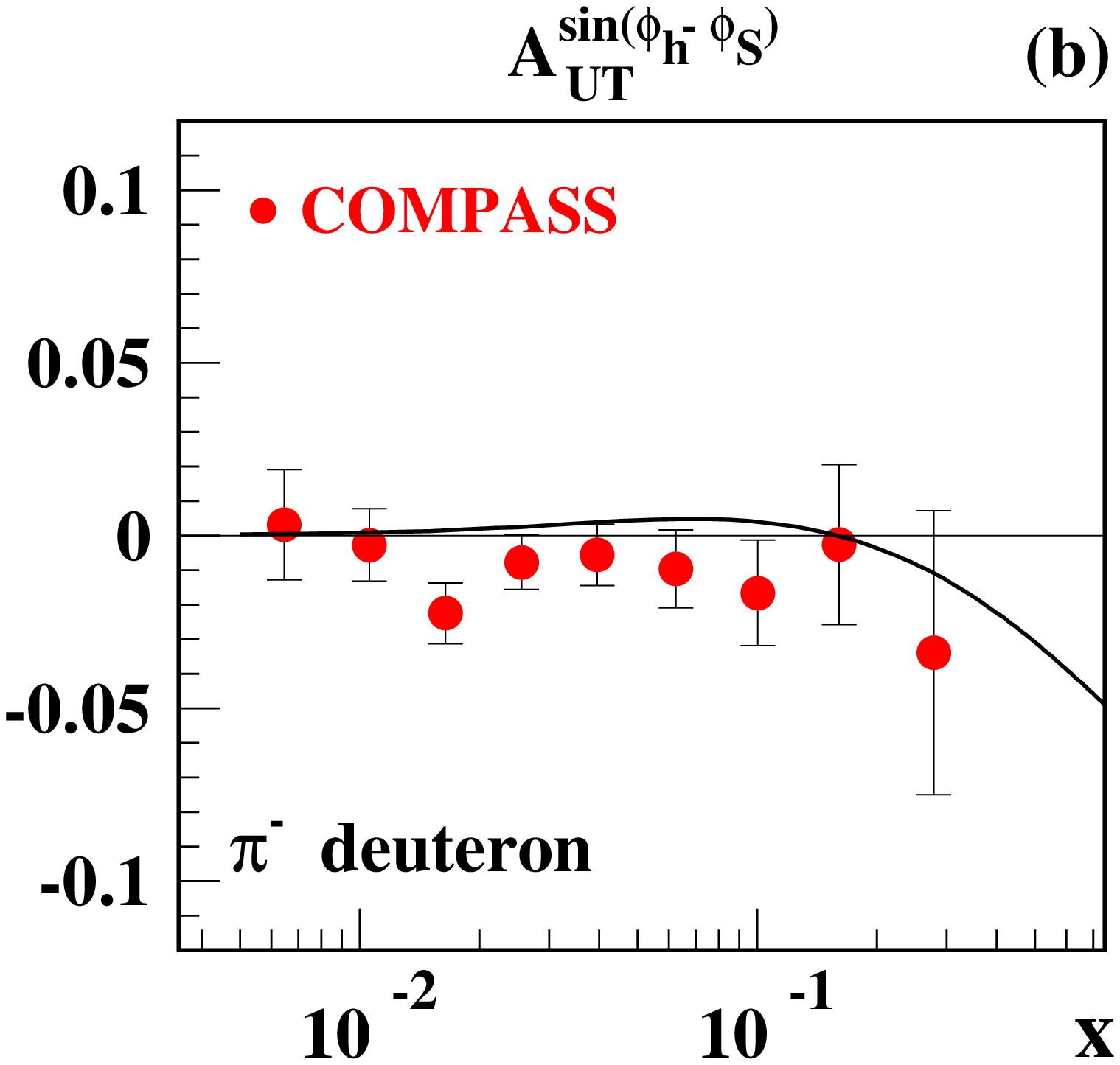}
\hspace{-12mm}
\includegraphics[height=5.2cm]{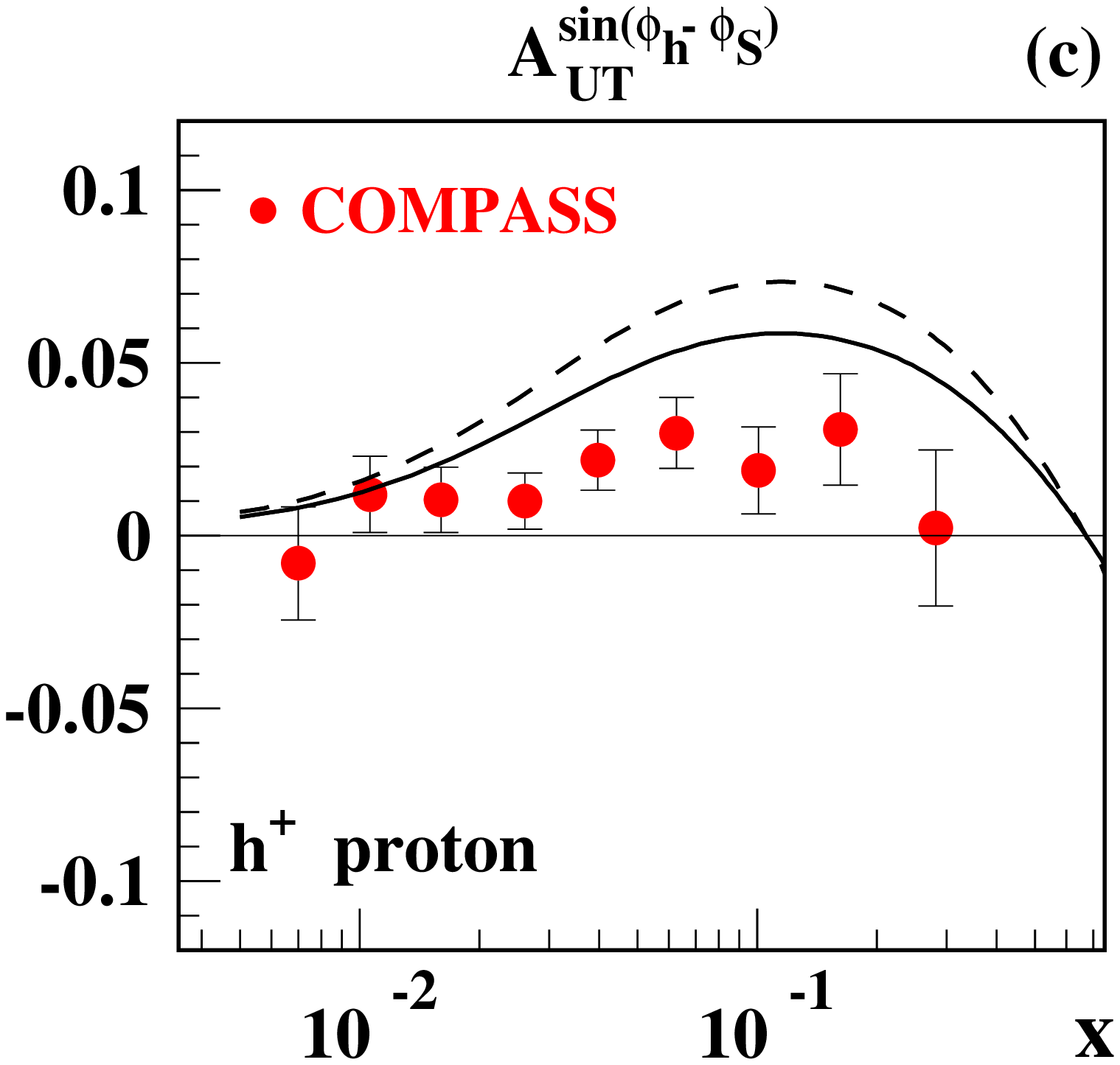}
\hspace{-12mm}
\includegraphics[height=5.2cm]{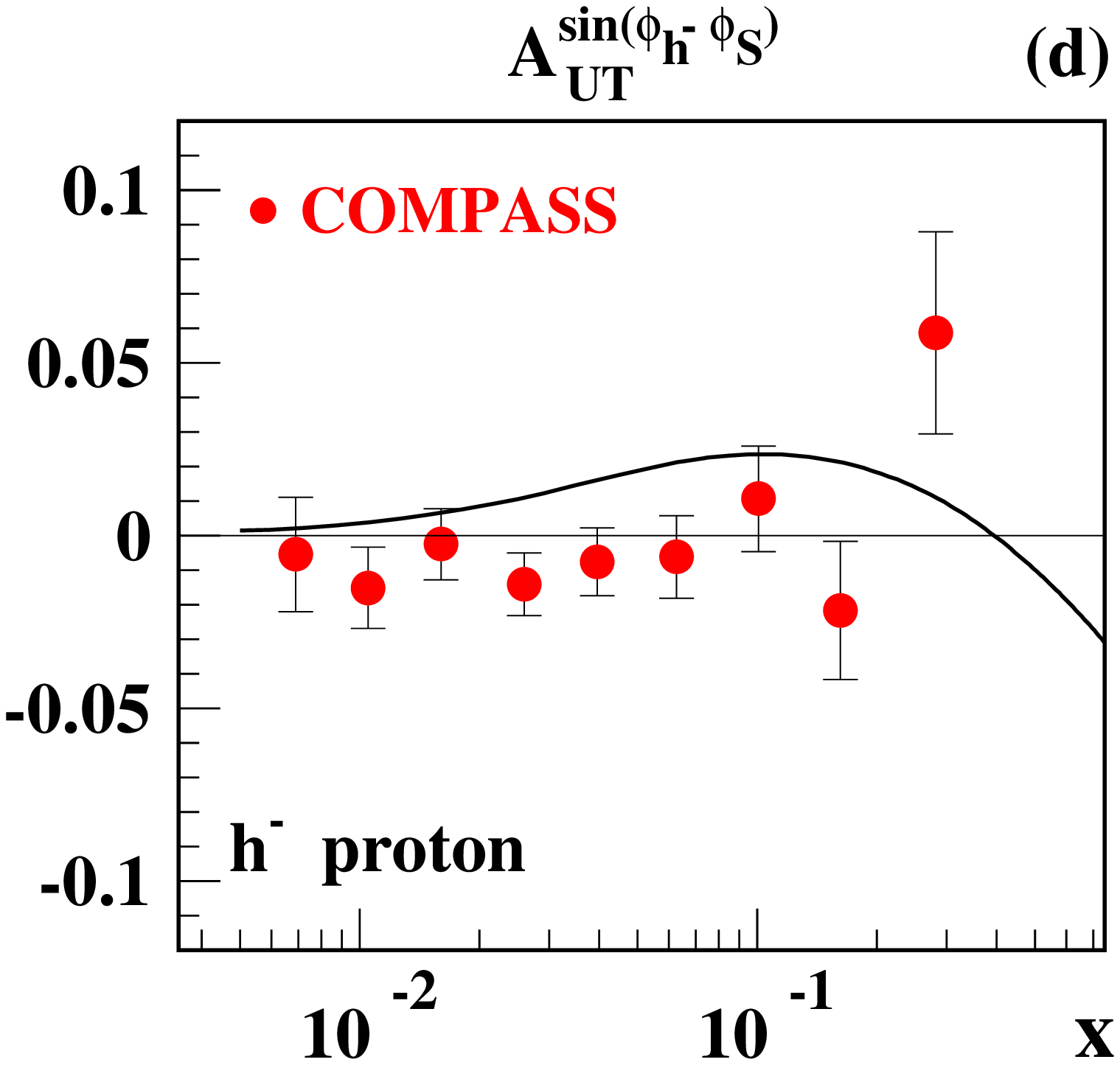}

\vspace{-9mm}

	\caption{\label{Fig-5:Sivers-COMPASS}
	$A_{UT}^{\sin(\phi_h-\phi_S)}$ for charged pion (hadron) production off 
        deuteron (proton) in SIDIS, as function of $x$. 	
        The COMPASS data are from \cite{Alekseev:2008dn,Alekseev:2010rw}.
        The theoretical curves are obtained on the basis of the 
        LCCQM predictions for $f_{1T}^{\perp(1)q}(x)$ \cite{Pasquini:2010af}.
        The dotted curve in panel (c) shows the results without the effects 
        of $p_T$-broadening (see text).}

  \includegraphics[height=5.2cm]{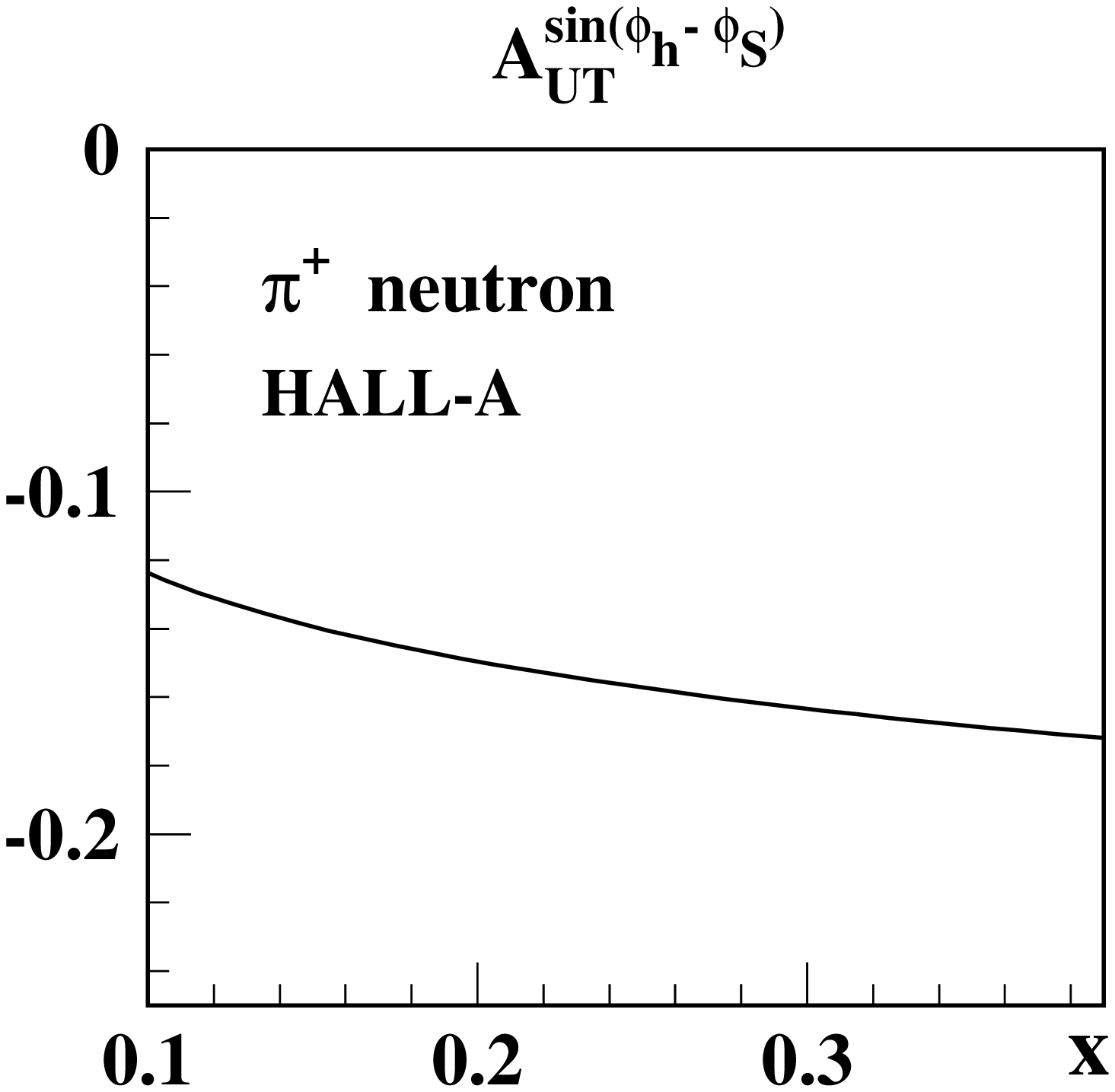}
\hspace{-8mm}
  \includegraphics[height=5.2cm]{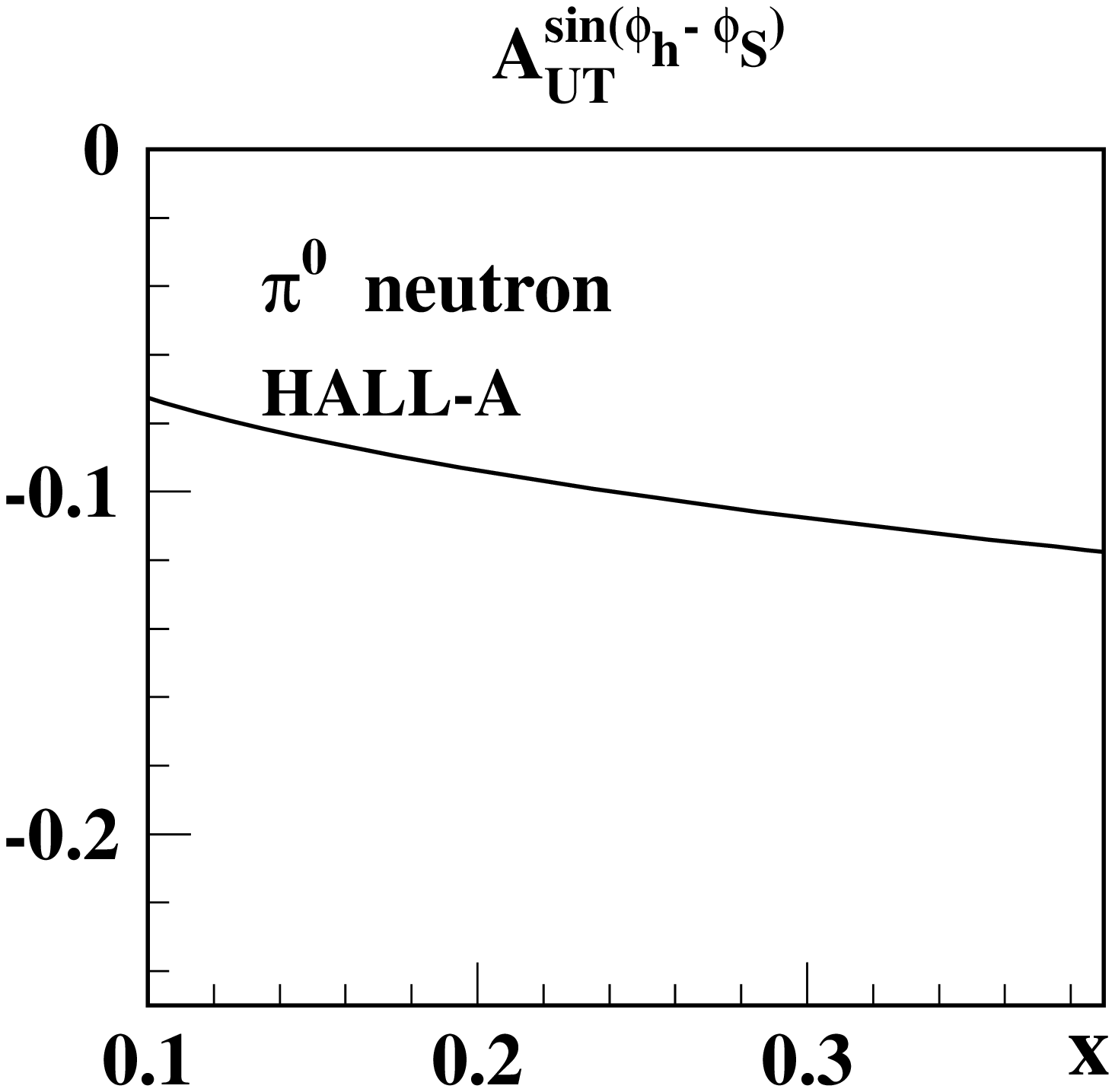}
\hspace{-8mm}
  \includegraphics[height=5.2cm]{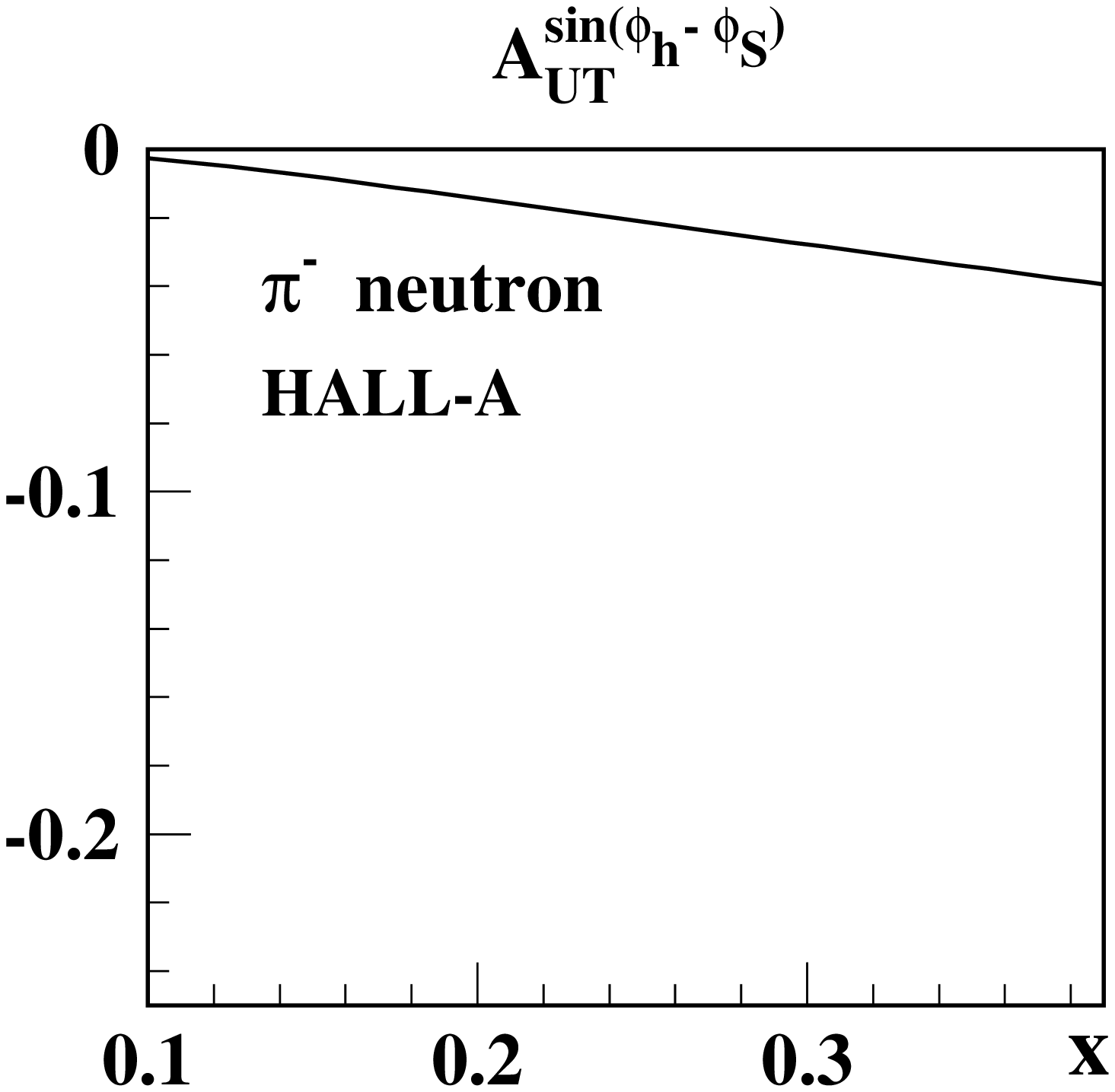}

\vspace{-10mm}

	\caption{\label{Fig-6:Sivers-Hall-A}
	$A_{UT}^{\sin(\phi_h-\phi_S)}$ for pion production off neutron in SIDIS, 
        as function of $x$.
        The theoretical predictions are obtained on the basis of results
        for $f_{1T}^{\perp(1)q}(x)$ from LCCQM \cite{Pasquini:2010af}
        for the kinematics of the Hall-A experiment at Jefferson Lab 
        \cite{Gao:2010av}.}

\end{figure*}
%

In this Section we discuss the asymmetry 
$A_{UT}^{\sin(\phi_h-\phi_S)}$ due to the Sivers effect.
With the model results referring to a low scale, it is not
only necessary to evolve the Sivers function in $x$,
but also to account for $p_T$-broadening \cite{Collins:1984kg}
(all this applies equally to the Boer-Mulders function). 
In principle, one could feed the Collins-Soper-Sterman (CSS) equation
\cite{Collins:1984kg} with the model results as initial conditions.
We refrain from this step, because the CSS formalism 
is not yet developed for cases including
polarization effects, while in the unpolarized case the result
is known: one would expect to reproduce, within model accuracy,
the phenomenological value $\la p_T^2(f_1)\ra\sim 0.4\,{\rm GeV}^2$ 
at HERMES energies. Therefore we proceed as follows.

We assume that the Gaussian shape is approximately preserved through
the CSS evolution but the Gaussian widths increase with energy
\cite{Schweitzer:2010tt,Aybat:2011zv}, and we assume that to lowest 
order approximation this $p_T$-broadening is polarization independent.
We use the model predictions for $f_{1T}^{\perp(1)q}(x)$ 
(approximately evolved in $x$ as discussed in Sec.~\ref{Sec-3:model})
which are presumably less affected by Sudakov effects \cite{Boer:2001he}
than $f_{1T}^{\perp q}(x)$ with which one also could work within the
Gauss model.
In the expression (\ref{Eq:GaussFUTSiv}) for the Sivers asymmetry
we use $\la K_T^2(D_1)\ra=0.16\,{\rm GeV}^2$
\cite{Schweitzer:2010tt} and the Gaussian width of the Sivers 
function, which is 
$\la p_T^2(f_{1T}^\perp)\ra \approx 0.9 \,\la p_T^2(f_1)\ra$
in the model \cite{Pasquini:2010af}. 
According to our assumption  one may expect this prediction
to be roughly valid also at experimentally relevant scales,
because to lowest order approximation $p_T$-broadening effects
are polarization-independent. 
In \cite{Boffi:2009sh} positive experience was made with 
such estimates of $p_T$-broadening effects.
Further studies are required for more precision.

In the numerator of the Sivers asymmetry we use 
model predictions for $f_1^a(x)$ LO-evolved to a scale of 
$2.5\,{\rm GeV}^2$.
For $D_1^a(z)$ we use the LO parameterization from \cite{Kretzer:2001pz} 
at the same scale. 

In this way we obtain in the kinematics of the HERMES 
experiment, $\la Q^2\ra\approx 2.5\,{\rm GeV}^2$
and $0.2< z < 0.7$, the results for the $x$-dependence of the
Sivers asymmetry for pion production off a proton target 
shown in Fig.~\ref{Fig-4:Sivers-HERMES}.
Keeping in mind that the quark model approach is not expected 
to be reliable in the small-$x$ region, we observe a good
description of the data within the accuracy of our approach.
(Due to the absence of sea quarks in our approach the 
Sivers asymmetries for kaons would be very similar to the
pion asymmetries. The explanation of a possible difference in 
$\pi^+$ and $K^+$ Sivers asymmetries \cite{Airapetian:2009ti}
is beyond the scope of our approach.)

\begin{figure*}[t!]

\vspace{-3mm}

\hspace{-2mm}
\includegraphics[height=5.2cm]{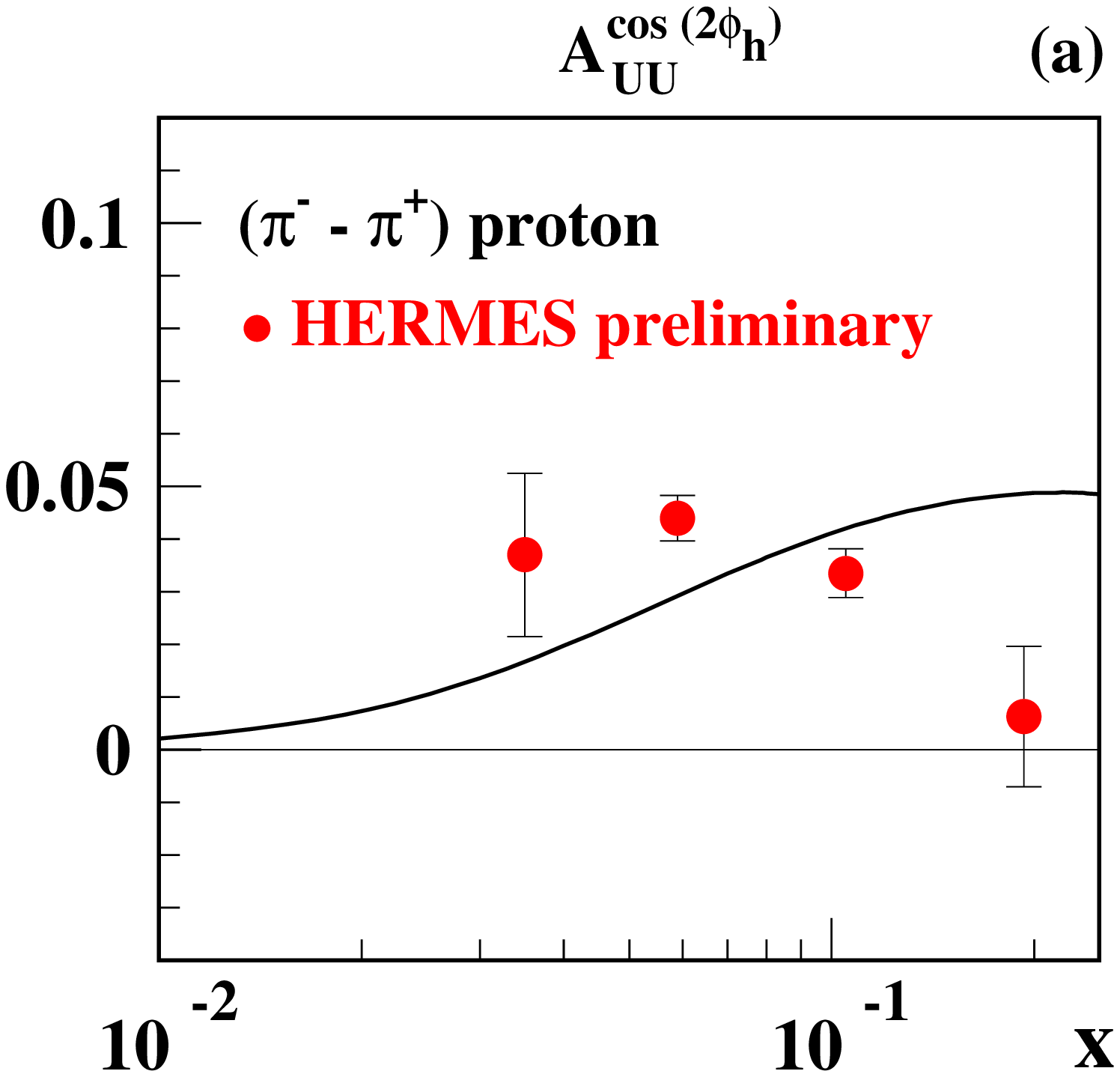}
\hspace{-12mm}
\includegraphics[height=5.2cm]{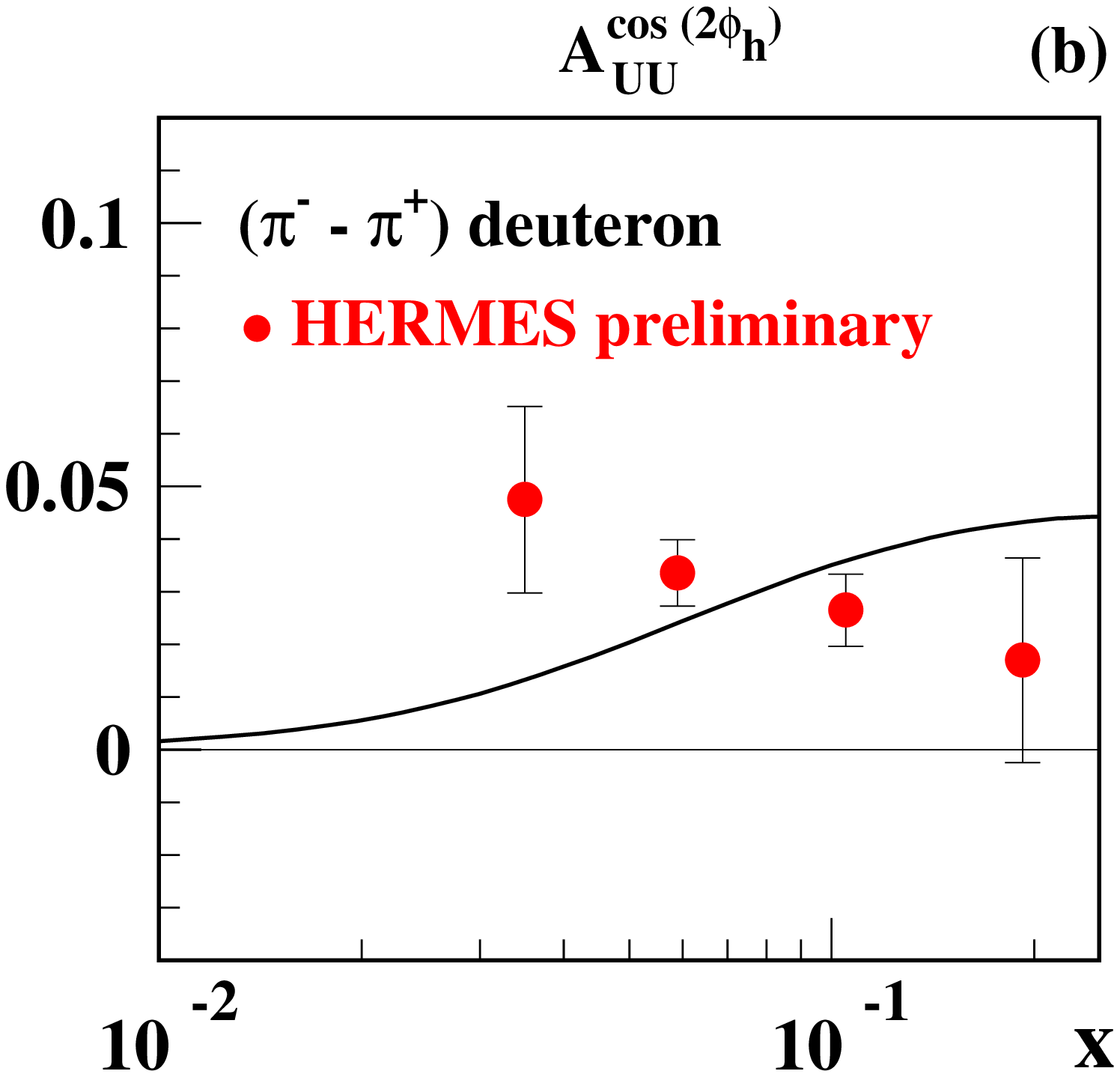}
\hspace{-12mm}
\includegraphics[height=5.2cm]{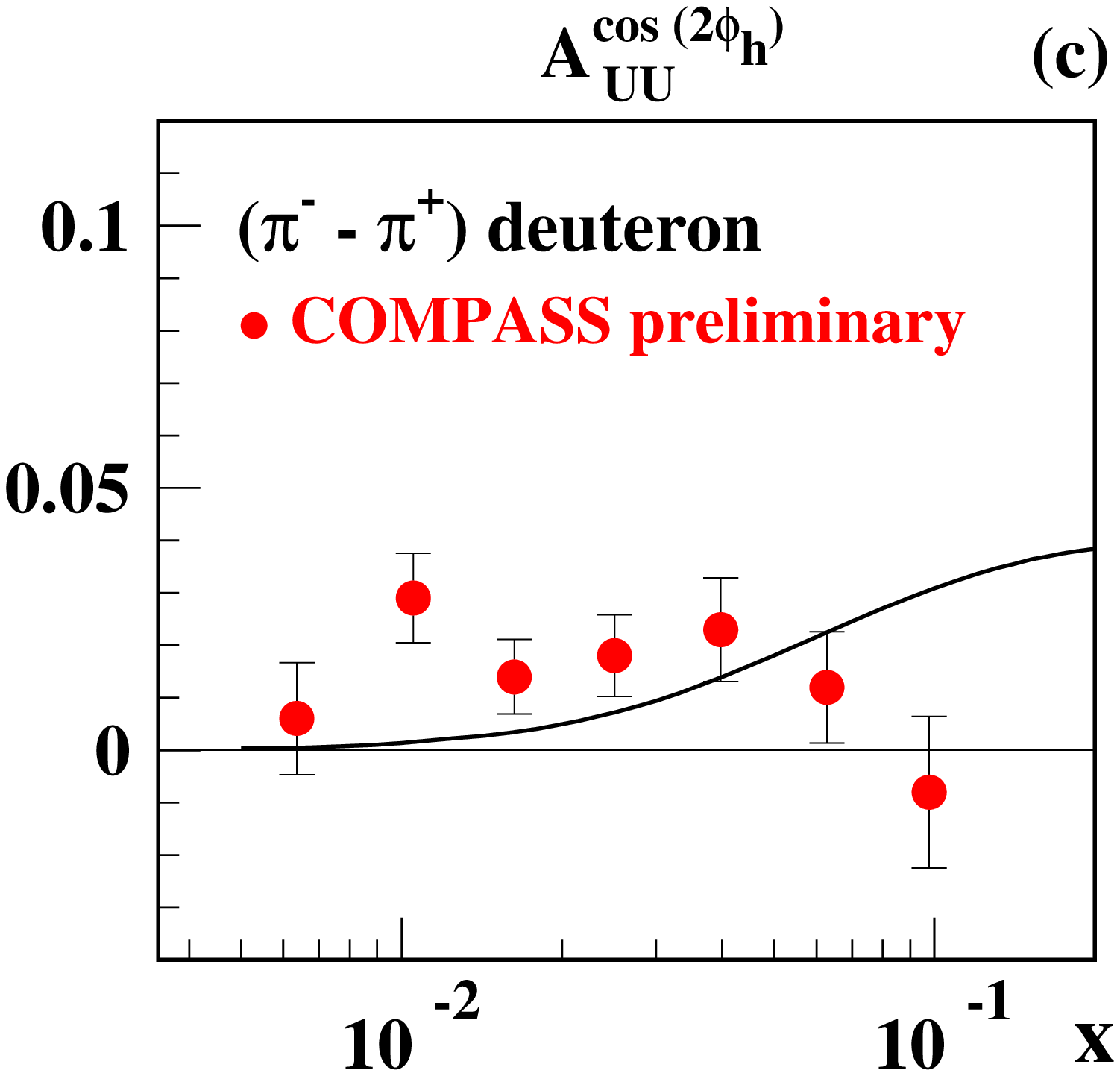}
\hspace{-12mm}
\includegraphics[height=5.2cm]{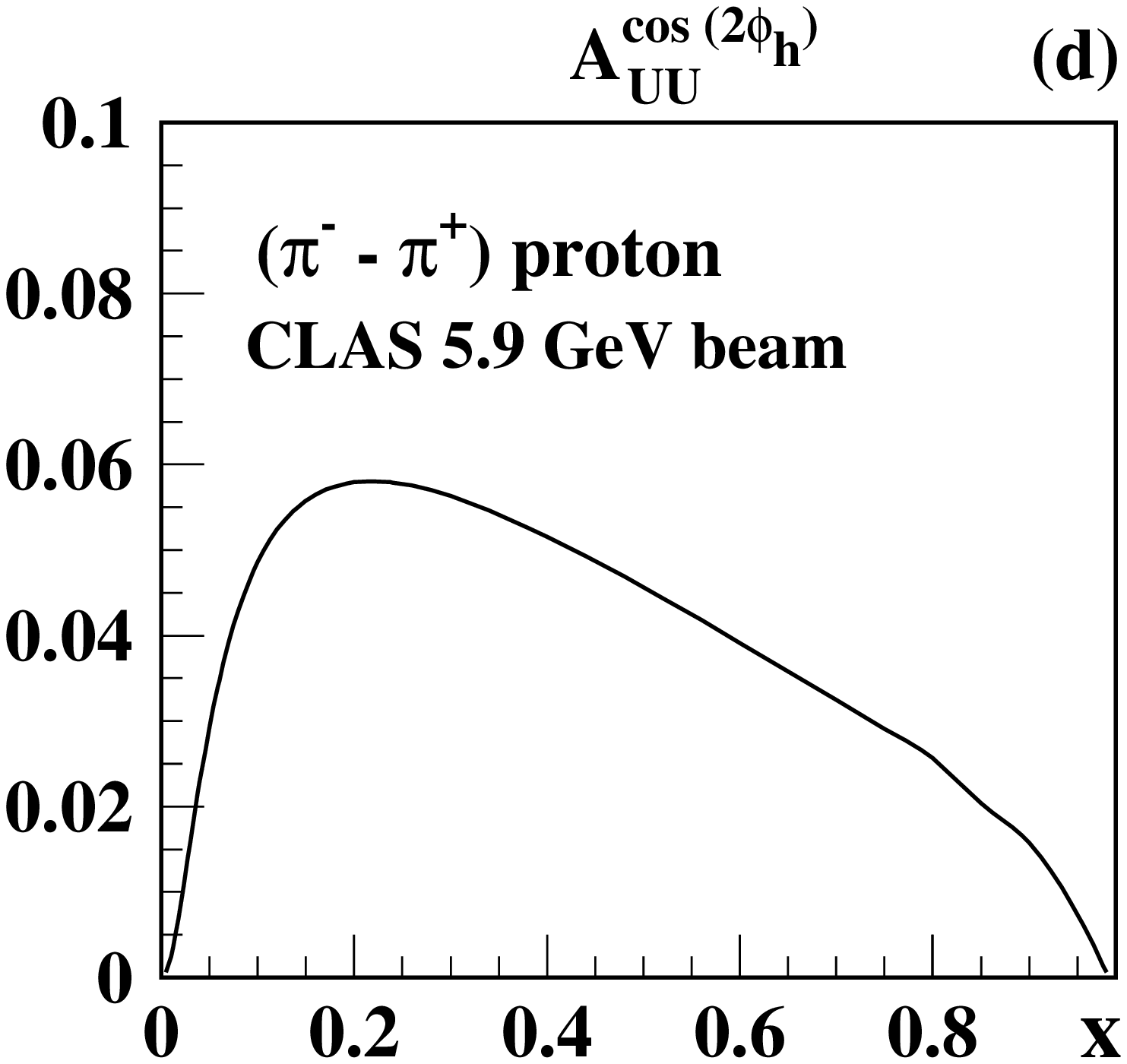}
\vspace{-8mm}

	\caption{\label{Fig-7:BM}
        The difference of azimuthal asymmetries $A_{UU}^{\cos(2\phi_h)}$ for 
        negative and positive pions or hadrons, as function of $x$. 
        The experimental points were obtained taking the differences of 
        preliminary $\pi^-$ and $\pi^+$ HERMES data \cite{Giordano:2010gq}, 
        and preliminary $h^-$ and $h^+$ COMPASS data \cite{Sbrizzai}.
        The error bars show the propagation of statistical errors,
        and do not include systematic errors.
        The theoretical curves are obtained using $h_1^{\perp(1)q}(x)$ 
        from the LCCQM \cite{Pasquini:2010af}.
        Panel (d) shows a prediction for Jefferson Lab.}
\end{figure*}
%

Next we discuss the COMPASS data which have $\la Q^2\ra$ similar to HERMES
and $0.2< z < 1$, 
but were measured at a significantly higher 
$s=(P+l)^2\approx 2M_NE_{\rm beam}$ with 
$E_{\rm beam}=160\,{\rm GeV}$ 
(vs.\ $E_{\rm beam}=27\,{\rm GeV}$ at HERMES).
From Drell-Yan it is known that at higher energies 
the TMDs tend to be broader
\cite{Schweitzer:2010tt} as follows from the CSS formalism 
\cite{Collins:1984kg,Aybat:2011zv}. There are indications 
that this is also the case in SIDIS \cite{Schweitzer:2010tt},
namely one has
$\la P_{h\perp}^2\ra=0.27\,{\rm GeV}^2$ at HERMES
\cite{Airapetian:2009jy} to be compared with
$\la P_{h\perp}^2\ra=0.41\,{\rm GeV}^2$ at COMPASS~\cite{Rajotte:2010ir}.
Both results refer to a common $z\sim 0.55$
but different $\la x\ra$.
If we assume an $x$-independent Gaussian width of $f_1^a$
(the data are compatible with this assumption \cite{Schweitzer:2010tt}),
the observed broadening of 
$\la P_{h\perp}^2\ra =\la K_T^2(D_1)\ra+z^2\la p_T^2(f_1)\ra$
has to be attributed to broadenings of the Gaussian widths of 
$f_1^a$ and $D_1^a$. In order to crudely estimate these effects 
we assume that at COMPASS the widths of $f_1^a$ and $D_1^a$
are equally broadened compared to HERMES as follows
\ba
&&    \la p_T^2(f_1)\ra_{\rm COMPASS} = \la p_T^2(f_1)\ra_{\rm HERMES} 
    + \delta\la\kappa_T^2\ra \;, \;\;\;\; \nonumber\\
&&    \la K_T^2(D_1)\ra_{\rm COMPASS} = \la K_T^2(f_1)\ra_{\rm HERMES} 
    + \delta\la\kappa_T^2\ra \;, \;\;\;\;\;\;\;\label{Eq:brodening}
\ea 
with $\delta\la\kappa_T^2\ra\approx 0.11\,{\rm GeV}^2$ needed
to explain the larger $\la P_{h\perp}^2\ra$ at COMPASS.
This change in parameters also affects (broadens)
the Gaussian width of the Sivers function (estimated here as
$\la p_T^2(f_{1T}^\perp)\ra \approx 0.9 \,\la p_T^2(f_1)\ra$,
see above).
In this way we obtain the results shown in Fig.~\ref{Fig-5:Sivers-COMPASS}.

We observe a good agreement of the model results with the COMPASS data on 
the Sivers effect in $\pi^\pm$ production from a deuteron target 
\cite{Alekseev:2008dn} in Figs.~\ref{Fig-5:Sivers-COMPASS}a,~b,
and charged hadron production from a proton target \cite{Alekseev:2010rw}
in Figs.~\ref{Fig-5:Sivers-COMPASS}c,~d.
(For simplicity we approximated the results for charged hadrons 
by charged pions, which account for about $90\,\%$ of charged hadrons 
at COMPASS energies.)

Notice that, had we neglected the $p_T$-broadening at COMPASS as
compared to HERMES, the model would have clearly overestimated $\pi^+$ 
data from proton at COMPASS, see Fig.~\ref{Fig-5:Sivers-COMPASS}c.
Indeed, Fig.~\ref{Fig-5:Sivers-COMPASS}c seems to indicate that the
$p_T$-broadening could even be somewhat stronger than estimated on 
the basis of (\ref{Eq:brodening}).
In Figs.~\ref{Fig-5:Sivers-COMPASS}a,~b,~d the neglect 
of broadening effects would be less significant, or completely
within error bars.

Finally, we present predictions for a forthcoming Hall-A experiment  
at Jefferson Lab \cite{Hall-A-neutron} with a $5.9\,{\rm GeV}$ beam.
Here $s$ is sufficiently close to HERMES, such that it is
not necessary to account for $p_T$-broadening effects, in contrast 
to COMPASS, Eq.~(\ref{Eq:brodening}) \cite{Schweitzer:2010tt}.
In this experiment $\la Q^2\ra\sim 2\,{\rm GeV}^2$ and 
$0.42<z<0.66$. The results for this kinematics are shown in 
Fig.~\ref{Fig-6:Sivers-Hall-A}. At Hall-A the asymmetry is larger 
because roughly $B_0(z)\propto z$ in (\ref{Eq:GaussFUTSiv}) 
and Hall-A probes larger $z$ compared to HERMES and COMPASS.

To summarize, we find that our model framework provides a satisfactory
description of the SIDIS data on the Sivers effect from HERMES and 
COMPASS \cite{Airapetian:2009ti,Alekseev:2008dn,Alekseev:2010rw}.
Future data from Jefferson Lab will allow further tests.

\section{\boldmath Boer-Mulders asymmetry}
\label{Sec-5:AUU}

In this section we focus on the asymmetry
$A_{UU}^{\cos(2\phi_h)}$ due to the Boer-Mulders effect.
The asymmetry (\ref{Eq:GaussFUUcos2phiBM}) is calculated
similarly to the Sivers asymmetry in Sec.~\ref{Sec-4:AUT}.

We use the model predictions for $h_1^{\perp(1)q}(x)$
from \cite{Pasquini:2010af}.
For the Collins function and its Gaussian width we use the 
information from \cite{Efremov:2006qm,Anselmino:2007fs},
and for the width of the Boer-Mulders function we use
the LCCQM model prediction
$\la p_T^2(h_1^\perp)\ra \approx 0.95 \,\la p_T^2(f_1)\ra$
\cite{Pasquini:2010af}, which we again assume to be 
approximately valid at experimentally relevant scales.
For $f_1^a(x)$ in the numerator of the asymmetry
we use the model predictions
LO-evolved to $2.5\,{\rm GeV}^2$~\cite{footnote-3}, and for $D_1^a(z)$ the LO 
parameterization from \cite{Kretzer:2001pz} at the same scale. 
The $p_T$-broadening effects at the higher energies in the
COMPASS as compared to HERMES, are estimated similarly
to Sec.~\ref{Sec-4:AUT} i.e.\ we use the HERMES value
$\la K_T^2(H_1^\perp)\ra\approx 0.25 \la K_T^2(D_1)\ra$ 
from \cite{Anselmino:2007fs},
and consider $p_T$-broadening analog to Eq.~(\ref{Eq:brodening}).

As discussed in detail in Sec.~\ref{Sec-2:SIDIS}, the Cahn-effect 
\cite{Cahn:1978se} generates an $1/Q^2$ power-correction 
to the $\cos(2\phi_h)$ modulation in the unpolarized SIDIS cross
section which cannot be neglected in the kinematics of the HERMES or 
COMPASS experiments \cite{Barone:2009hw,Schweitzer:2010tt}. 
In principle, one can try to model this power-correction 
\cite{Barone:2009hw}. 
For that one could use the updated phenomenological results for 
$\la p_T^2(f_1)\ra$ and $\la K_T^2(D_1)\ra$ \cite{Schweitzer:2010tt} 
which are sufficient to determine the Cahn-effect.
Alternatively, one could explore data on the $\cos(2\phi_h)$-asymmetry of 
neutral pions, where due to the flavor-dependence of the Collins function
\cite{Efremov:2006qm,Anselmino:2007fs} the leading-twist Boer-Mulders 
effect largely cancels, see Appendix.
In any case, this step constitutes an {\sl additional }
modeling step (of a non-factorizable twist-4 contribution),
and we prefer to avoid it.

However, the Cahn effect contamination in $A_{UU}^{\cos(2\phi_h)}$
is flavor-independent to a good approximation, and largely cancels out 
in differences of $\pi^-$ and $\pi^+$ 
asymmetries. Such differences can be determined 
from the preliminary data \cite{Giordano:2010gq,Sbrizzai}
and we shall confront our model results with them instead.
The results are shown in Figs.~\ref{Fig-7:BM}a-c.
(We again approximate at COMPASS $h^\pm$ results by $\pi^\pm$.)

Again it is important to recall that the quark results 
should not be expected to be reliable in the small-$x$ region. 
In the region of $x\sim 0.1$ the model describes well the size of
the asymmetry differences, but it seems not to follow the trend of 
the data at larger $x\gtrsim 0.2$. However, one has to keep in mind
the preliminary status of the data \cite{Giordano:2010gq,Sbrizzai}.
Moreover, considering systematic errors of the data
which are not included in Figs.~\ref{Fig-7:BM}a-c, 
the discrepancy could be well within model accuracy.

From the available Jefferson Lab
data \cite{Mkrtchyan:2007sr,Osipenko:2008rv} 
the difference of $\pi^-$ and $\pi^+$ $\cos(2\phi_h)$-asymmetries 
cannot be accessed, but forthcoming CLAS data will allow that
\cite{Gohn:2009zz}. 
In the CLAS experiment we have $\la Q^2\ra \sim 1.9\,{\rm GeV}^2$ 
with $0.1 < x < 0.6$ and $0.4 < z < 0.7$. The predictions for 
this kinematics are shown in Fig.~\ref{Fig-7:BM}d. 
Since the $x$ values of the CLAS experiment cover the  region 
dominated by the valence-quark contribution, these data will 
provide an important test of the model.

To summarize, our approach is compatible with preliminary data
from COMPASS and HERMES. 
Further insights can be expected from Jefferson Lab, before
\cite{Gohn:2009zz} and after the 12 GeV beam energy upgrade 
\cite{Avakian:LOI}, and on long term from the future 
Electron-Ion Collider \cite{Anselmino:2011ay}.


\section{Conclusions}
\label{sect:conclusions}

In this work we have studied the leading-twist azimuthal asymmetries 
in SIDIS due to T-odd TMDs on the basis of predictions from the
light-cone constituent-quark model \cite{Pasquini:2010af}.
Since the model results refer at a  low hadronic scale, we discussed 
 how to take into account the effects of the evolution for the description 
of data referring to high scales of typically several ${\rm GeV}^2$.
We tackled this issue in two steps.
First, for the $p_T$-dependence of the distributions we employed the 
Gaussian Ansatz and expressed the asymmetries in terms of 
(1)-transverse moments of TMDs. 
The Gaussian widths of the distributions were assumed $x$-independent, 
which is supported by phenomenology, and $p_T$-broadening effects were 
estimated.
In the second step, we evolved the transverse moments of TMDs to  
experimental scales by employing those evolution equations which 
seem most promising to simulate the correct evolution. 
For the Sivers distribution we used the non-singlet 
evolution pattern of $f_1^a(x)$. This allows to preserve
the Burkardt sum rule ---  valid at the initial scale of the model
\cite{Pasquini:2010af} --- also at higher scales.
For the chiral-odd Boer-Mulders function we used the evolution pattern 
of transversity.

We obtained a good description of the Sivers asymmetry, and 
satisfactory results for the Boer-Mulders asymmetry in comparison 
with available experimental data from HERMES and COMPASS.
In the case of the Boer-Mulders asymmetry we considered differences of
asymmetries for $\pi^-$ and $\pi^+$ to avoid the modeling of
twist-4 power corrections ('Cahn effect').
Furthermore, we presented model predictions for forthcoming experiments 
at Jefferson Lab, 
which will extend the available data far into the valence-$x$ region
where the model is expected to work best, and provide
an important test of its dynamics.

Our results indicate that the use of the one-gluon-exchange-mechanism
to model T-odd TMDs (as implemented in \cite{Pasquini:2010af})
yields phenemenologically reasonable results, although a truncation
of the expansion of the gauge-link at ${\cal O}(\alpha_s)$ seems
not a priori justifiable. 

The present work completes the study in Ref.~\cite{Boffi:2009sh} where 
leading-twist spin asymmetries due to T-even TMDs were calculated.
We observe that the light-cone constituent-quark model,
based on overlap representation of TMDs in terms of light-cone wave 
functions, provides a 
good description of intrinsic transverse parton momentum effects in the 
range of applicability of the model.

\vspace{0.2cm}

\noindent{\bf Acknowledgements.}
We thank M.~Contalbrigo from the HERMES collaboration, 
and E.~Kabuss, A.~Martin and G.~Sbrizzai from the COMPASS collaboration
for making available the (final and preliminary) data. 
We also acknowledge helpful discussions with A.~V.~Efremov, W.~Gohn, 
J.-F.~Rajotte, T. Rogers and F.~Yuan.
B.~P.~is grateful for the hospitality to the Department of Physics of the 
University of Connecticut where this work was initiated.
The work was supported in part by DOE contract DE-AC05-06OR23177, under
which Jefferson Science Associates, LLC,  operates the Jefferson Lab,
by 
the Research Infrastructure Integrating Activity
``Study of Strongly Interacting Matter'' (acronym HadronPhysics2, Grant
Agreement n. 227431) under the Seventh Framework Programme of the
European Community, and by the Italian MIUR through the PRIN 
2008EKLACK ``Structure of the nucleon: transverse momentum, transverse 
spin and orbital angular momentum''.

\appendix
\section{Remark on Cahn effect}

Finally, a remark is in order on $A_{UU}^{\cos(2\phi_h)}$ for
definite pions, where the Cahn effect $1/Q^2$-correction
cannot be neglected \cite{Barone:2009hw,Schweitzer:2010tt}
though it could be small in largest $x$-bins at COMPASS.
In principle, one may model this contribution \cite{Barone:2009hw}
with due care to the energy dependence of the Gauss parameters
\cite{Schweitzer:2010tt,Aybat:2011zv}.
However the problem remains how to test independently such an
additional modeling of a presumably non-factorizing twist-4 term.
We are not aware of a 'rigorous procedure', but the following
observation may turn out helpful.
The favored ($u\to\pi^+$) and unfavored ($u\to\pi^-$) Collins
fragmentation functions are similar in magnitude but have opposite
signs \cite{Efremov:2006qm,Anselmino:2007fs}.

In the $A_{UT}^{\sin(\phi_h+\phi_S)}$ asymmetry for $\pi^0$
(which is due to the Collins effect and potentially not or less
affected by power-corrections) this yields nearly exact
flavor cancellations, as seen in data \cite{Airapetian:2010ds}.
Since this is a property of the Collins function, one
expects the Collins effect to also cancel out in the
$\pi^0$ $\cos(2\phi_h)$-asymmetry, i.e.\
\be
     A_{UU}^{\cos(2\phi_h)}(\pi^0) \approx
     A_{UU,\,\rm Cahn}^{\cos(2\phi_h)} < 0\;,
     \label{Eq:BM-Cahn-pi0}
\ee
where $A_{UU,\,\rm Cahn}^{\cos(2\phi_h)}$ is the Cahn effect
contribution, which is negative \cite{Cahn:1978se} and
largely flavor independent \cite{Barone:2009hw,Schweitzer:2010tt}.
This would then mean that
\ba
     A_{UU}^{\cos(2\phi_h)}(\pi^+) \approx
     A_{UU}^{\cos(2\phi_h)}(\pi^+)_{\rm BM} +
     A_{UU,\,\rm Cahn}^{\cos(2\phi_h)}\, ,
     \label{Eq:BM-Cahn-pi+}\\
     A_{UU}^{\cos(2\phi_h)}(\pi^-) \approx
     A_{UU}^{\cos(2\phi_h)}(\pi^-)_{\rm BM} +
     A_{UU,\,\rm Cahn}^{\cos(2\phi_h)}\, .
     \label{Eq:BM-Cahn-pi-}
\ea
\ \\
This is not to be confused with relations due to
isospin invariance, which allow one to express
$\pi^0$ cross sections in terms of $\pi^{\pm}$
cross sections (i.e.\ which connect the
numerators {\sl or} the denominators of the asymmetries).
Eqs.~(\ref{Eq:BM-Cahn-pi0})-(\ref{Eq:BM-Cahn-pi-})
indicate how to model the Cahn-effect in a given experiment.
This might be a more reliable
procedure than using other sources of information.
The procedure can be iteratively improved to
take into account flavor-dependencies in the Cahn-effect
and non-exact cancellations of the Collins effect
in the neutral pion $\cos(2\phi_h)$-asymmetry.
In particular,
Eqs.~(\ref{Eq:BM-Cahn-pi+}) and (\ref{Eq:BM-Cahn-pi-})
show our underlying assumption in Sec.~\ref{Sec-5:AUU} that
in the difference of charged pion $\cos(2\phi_h)$-asymmetries
the Cahn effect largely cancels out.

\end{document}